\documentclass[preprint]{aastex}

\usepackage{natbib}

\newcommand{\ie}{{\em i.e.\ }}
\newcommand{\eg}{{\em e.g.,\ }}
\newcommand{\etal}{{\em et al.\ }}
\newcommand{\etc}{{\em etc.}}

\newcommand{\BVRI}{{$B, V, R, \&\ I $}}
\newcommand{\BRIK}{{$B, R, I, \&\ K^{'}$}}

\slugcomment{Accepted to AJ}

\shorttitle{Galaxy Cluster Membership}
\shortauthors{Brunner & Lubin}

\begin{document}

\title{A Probabilistic Quantification of Galaxy Cluster Membership}

\author{R.J. Brunner and L.M. Lubin\altaffilmark{1}} \affil{Department
of Astronomy, California Institute of Technology,\\ MC 105-24,
Pasadena, CA 91125}
\email{rb@astro.caltech.edu}

\altaffiltext{1}{Hubble Fellow}

\begin{abstract}

Clusters of galaxies are important laboratories for understanding both
galaxy evolution and constraining cosmological quantities. Any
analysis of clusters, however, is best done when one can reliably
determine which galaxies are members of the cluster. While this would
ideally be done spectroscopically, the difficulty in acquiring a
complete sample of spectroscopic redshifts becomes rather daunting,
especially at high redshift where the background contamination becomes
increasingly larger. Traditionally, an alternative approach of
applying a statistical background correction has been utilized, which,
while useful in a global sense, does not provide information for
specific galaxies. In this paper, we develop a more robust technique
which uses photometrically estimated redshifts to determine cluster
membership. This technique can either be used as an improvement over
the commonly used statistical correction method or it can be used to
determine cluster candidates on an individual galaxy basis. By tuning
the parameters of our algorithm, we can selectively maximize our
completeness or, alternatively, minimize our contamination.
Furthermore, our technique provides a statistical quantification of
both our resulting completeness and contamination from foreground and
background galaxies.

\end{abstract}


\section{Introduction}

In the wake of the Hubble deep field campaigns~\citep{williams96}, and
with the growth of large photometric surveys (\eg DPOSS, SDSS, \etc),
the techniques involved in calculating photometric redshifts have
become increasingly more sophisticated and precise~\citep[see][for a
recent synopsis of the field]{weymann99}. Accuracies of $\sigma_{z} =
0.05$ can be routinely achieved~\citep{brunner99}. Therefore, it is
now possible to use photometric redshifts to answer specific
scientific questions~\citep[\eg][]{connolly98,brunner00}. In light of
this, we have begun a program to use accurate photometric redshifts to
study the galaxy populations in high-redshift clusters of galaxies.
For this study, we have examined the most well-studied sample of
optically-selected, high-redshift clusters to date -- that of Oke,
Postman \& Lubin (1998). Over the past few years, Oke \etal (1998)
have undertaken a detailed photometric, spectroscopic, and
morphological survey of nine candidate clusters at $z \ga 0.6$.  The
survey consists of deep $BVRIK^{'}$ photometry, over 130 low-resolution
Keck spectra per cluster field, and high-spatial-resolution WFPC2
imagery from the {\it Hubble Space Telescope} (HST).

Previous work on similar galaxy cluster samples at various redshifts
has relied on statistical corrections in order to remove contamination
due to field galaxies when spectroscopic identifications are not
available for all galaxies in the cluster
field~\citep[\eg][]{aragon93,lubin96,smail97,dressler97,couch98,stanford98,lubin98a}.
These types of corrections, however, are fraught with potential
systematic effects, including cosmological variance, weak lensing, and
sample selection criteria. While these effects will only serve to
increase the noise of any measurement made over a statistically
complete sample, they will systematically affect any measurements made
on an individual cluster (\eg the Morphological fraction or the
Morphology-Density relationship), introducing a bias into the actual
measurement process. Furthermore, these line-of-sight contamination
corrections apply to the entire galaxy sample and do not provide any
information for specific galaxies. As a result, we have decided to
explore the role photometric redshifts could play in improving the
reliability of identifying galaxy cluster
members~\citep{kodama99,lubin-ociw99}.

In this paper, we first describe in Section~\ref{data-sec} the details
of the data observations, reduction, and catalog generation,
emphasizing relevant points which are important for the estimation of
photometric redshifts. In Section~\ref{photoz-sec} we describe the
details of our redshift estimation technique. Next, we detail the
cluster galaxy identification procedure in
Section~\ref{pzcluster-sec}. We conclude with a discussion of our
results and future applications.

\section{Data~\label{data-sec}}

For the photometric redshift analysis, we have used the photometric
and spectroscopic data from the original cluster survey of Oke \etal
(1998).  This survey consists of nine candidate clusters at redshifts
of $z \ga 0.6$ : Cl 0023+0423, Cl 0231+0048, Cl 0943+4804, Cl
1324+3011, Cl 1325+3009, Cl 1604+4304, Cl 1604+4321, Cl 1607+4109, and
Cl 2157+0347.  Oke \etal (1998) have determined that six of the nine
candidate clusters are indeed real density enhancements, consistent
with that of a cluster of galaxies.  The observational data relevant
for the work described in this paper consists of $B$, $V$, $R$, $I$,
and $K^{'}$ band imaging and spectroscopic observations of selected
galaxies in each field.

\subsection{Observations and Data Reduction}

All of the ground-based optical observations, both broad band and
spectroscopic, were taken with the Low Resolution Imaging Spectrometer
(LRIS; Oke \etal 1995) on either the Keck I or the Keck II 10-m
telescopes. The infrared imaging was performed with the Infrared
Imaging Camera (IRIM) on the Mayall 4-m telescope at Kitt Peak
National Observatory. In this section, we briefly describe the
observations and the data reduction; however, the reader is referred
to Oke \etal (1998) for a complete account of these observations.

\subsubsection{Broad-Band Optical}

The photometric survey was conducted in four broad band filters,
$BVRI$, which closely match the Cousins system. The response curves of
these filters are shown in Figure 1 of Oke \etal (1998). In imaging
mode, LRIS covers a field-of-view of $6 \times 8$ arcmin. The total
exposure times in each filter are 3600, 2000, 1200, and 900 seconds
for $B$, $V$, $R$, and $I$, respectively, and were chosen to give
fairly uniform errors in the photometry.  The $B$, $V$, and $R$
observations consisted of two equal exposures to allow for accurate
cosmic ray rejection.  The $I$ exposure time was broken into 3 equal
exposures in order to cope with the cosmic rays, as well as to avoid
approaching the CCD saturation level.

The LRIS imaging data were reduced in the standard fashion.  All
frames were bias-subtracted, and pixel-to-pixel sensitivity variations
were removed using dome flats. Large-scale gradients were removed by
dividing each frame by a normalized two-dimensional spline fit to the
sky values in a sky-flat.  The sky-flat was created by generating a
median image from a stack of frames for each night and passband.  For
the $I$ band data, a two-dimensional fringe map was also created from
the median filtered image by removing large-scale gradients within the
median image. Fringing was removed from each $I$ band frame by
subtracting a suitably scaled version of the fringe map.

Image registration for a given cluster field was performed by
identifying approximately 10 unsaturated stars (detectable in all 4
passbands) to be used as astrometric reference points. The mean $X$
and $Y$ offsets of these stars in every frame taken of the cluster
were computed relative to their locations in a fiducial $B$ band
image. All image data for the cluster were shifted to match the
$B$ band coordinate system using a flux conserving Lagrangian
interpolation scheme to achieve registration at the sub-pixel level.
Once all frames of a given cluster were registered to a common
coordinate system, the independent exposures in each passband were
co-added to produce the final four $BVRI$ images.

The Keck imaging has been calibrated to the standard
Cousins-Bessell-Landolt (Cape) system through exposures of a number of
Landolt standard star fields (Landolt 1992).  For a circular aperture
with a radius of $3^{''}$, the approximate limiting magnitudes are $B
= 25.1$, $V = 24.1$, $R = 23.5$, and $I = 21.7$ for a 5-$\sigma$
detection.

\subsubsection{Spectroscopic}

Multi-slit observations of the cluster field were made with LRIS in
spectroscopic mode using a $300~{\rm g~mm^{-1}}$ grating blazed at
5000 \AA. The chosen grating provided a dispersion of 2.35 \AA\ per
pixel and a spectral coverage of 5100 \AA. The grating angle was set
in order to provide coverage from approximately 4400 \AA\ to 9500 \AA\
in the first order. A GG495 glass filter was used to eliminate the
overlapping second order spectrum; there is, therefore, no second
order contamination below 9700 \AA.  In order to obtain the full
wavelength range along the dispersion axis, the field-of-view of the
spectral observations was reduced from that of the imaging mode to
approximately $2 \times 8$ arcmin.

Spectroscopic candidates were chosen from preliminary $R$ band imaging
of each cluster field. All objects brighter than approximately $R =
23.5 \pm 0.1$ within the spectroscopic field-of-view of LRIS were
included as candidates for slit-mask spectra.  For each cluster field,
six different slitmasks were made with approximately 30 objects per
mask (including guide stars and duplicate observations). The exposure
time for each mask was 1 hour. Flat-fielding and wavelength
calibration were performed using internal flat-field and arc lamp
exposures which were taken after each science exposure.

In practice, about 130 spectra were acquired per cluster field, with
$\sim 100$ yielding measurable redshifts. A redshift measurement
procedure which relies, in part, on visual inspection was used. The
quality of the redshift identification was ranked with a number from 1
to 4, which roughly corresponds to the number of features used to
identify the redshift.  A quality of 4 means that the redshift is
certain; a quality of 1 means that only one emission line was
observed, and the redshift is only possible.  Because the
fore/background contamination rate is so large at our cluster
redshifts, approximately 50 -- 85\% of the galaxies turn out to be
field galaxies, rather than cluster members (Oke \etal 1998; Postman
\etal 1998, 2000).

\subsubsection{Infrared}

Deep infrared imaging of the cluster sample was taken with IRIM which
contains a $256 \times 256$ NICMOS3 HgCdTe array and has a resolution
of 0.60 arcsecond~per~pixel on the 4-m telescope. The field-of-view
($154 \times 154$ arcseconds) covers the central region of each
cluster. In order to make the deepest possible observations over this
region the entire LRIS field-of-view was not mosaiced.  Observations
were made in the $2.2\mu\ K^{'}$ band. Each central field was observed
using a $4 \times 4$ dither pattern with a stepsize of $10^{''}$ and a
total extent of $30^{''}$. Each exposure had an effective exposure
time of 1 minute, with 4 co-additions of individual,
background--limited 15 second integrations. The total integration time
on an individual cluster varied between 3 and 4.4 hours. Because the
fields were not excessively crowded, we were able to use in-field
dithering to create a global sky flat.

The $K^{'}$ cluster data were reduced using the Deep Infrared
Mosaicing Software (DIMSUM), a publicly available package of IRAF
scripts. This software generates not only a final stacked image but
also a corresponding exposure image, where each pixel encodes the
total number of seconds in the corresponding stacked image. This
exposure image was appropriately scaled and used as a weight image by
SExtractor during the catalog generation (see \S 2.2).

The data were linearized, trimmed to exclude masked columns and rows
on the edges of the arrays, and dark--subtracted using dark frames of
the same exposure length as the observations.  All images of a given
night were flattened by a super flat made from a series of dome flats
taken during the previous day. As part of the DIMSUM procedure, sky
subtraction was done by subtracting a scaled median of nine temporally
adjacent exposures for each frame.  A first-pass reduction was used to
create an object mask for each frame.  This mask was created from a
fully stacked mosaic image. It therefore excluded not only the bright
objects, but also those objects too faint to be detected in an
individual exposure. In the second pass, the object mask was used to
avoid object contamination of the sky flat in the production of sky
frames. Final mosaicing of the images of each cluster were made with a
replication of each pixel by a factor of 4 in both dimensions. This
procedure conserves flux while eliminating the need for interpolation
when the individual frames are co-aligned. A bad pixel mask was used
to exclude bad pixels from the final summed images.

Absolute photometric transformations were derived from the
observations of Persson \etal (1998) HST standard stars. Each standard
star was observed every night in five separate array positions. The
observations of the standard stars were reduced in a similar manner to
the cluster data (see Oke \etal 1998). An approximate limiting
magnitude of $K^{'} = 20$ for a $5 \sigma$ detection was reached in a
standard aperture of radius $3\farcs{0}$.
 
\subsection{Catalog Generation}

We used SExtractor version 2.1.5~\citep{bertin96} to perform the
source detection and photometry because this program is able to detect
objects in one image and analyze the corresponding pixels in a
separate image. Applied uniformly to multi-band image data (i.e. use
the same detection image for all measurement images), this method will
produce a matched aperture photometric catalog. We generated an
optimal detection image from the \BVRI\ images using a $\chi^2$
process~\citep{szalay98}. Briefly, this process involves convolving
each input image with a Gaussian kernel matched to the seeing. The
convolved images were squared and normalized so that they have zero
mean and unit variance. The four processed images (corresponding to
the original \BVRI\ images) were coadded, forming the $\chi^2$
detection image.

In order to determine the optimal threshold parameters for source
detection, we compared a histogram of the pixel distribution in the
$\chi^2$ image with a $\chi^2$ function with four degrees of freedom.
(This distribution theoretically corresponds to the background pixel
distribution for our coadded $\chi^2$ image.) By taking the difference
between the two histograms (pixel minus theory), we generated the
histogram of pixel values which were due to the objects that we were
trying to detect. We defined the Bayesian detection threshold as the
intersection of the sky (or theoretical prediction) and object pixel
distributions (\ie where the object pixel flux becomes dominant). To
convert this empirical threshold for use with SExtractor, we scaled the
$\chi^2$ threshold (which is a flux per pixel value) into a surface
brightness threshold (which is in magnitudes per square arcsecond).

The infrared images were cataloged separately due to their radically
different field-of-view and pixel scale. The catalog was generated
using SExtractor as before but in single image mode (\ie the detection
image was also the measurement image). In order to maximize the number
of detections, we utilized the exposure image generated by DIMSUM
during the data reduction (see \S 2.1.3) to provide a weight map in
order to properly determine object parameters in an image with varying
signal-to-noise (due to the different total integration times in
different pixels).

All of our resultant analyses use the total magnitudes calculated by
SExtractor. The optical magnitudes were, by design, matched apertures
which reduces the scatter in any photometric redshift technique. Since
the $K^{'}$ band images were not co-registered with the optical images, the
aperture can be slightly different. However, since the total apertures
were rather large in angular extant, any ``aperture effects'' on our
resulting analysis should be small ($\lesssim 0.02^m$) compared to the
more dominant effects of photometric errors.

\subsection{Cross-identifications}

The last step in preparing these data for analysis was to combine the
optical imaging, near-infrared imaging, and the spectroscopic
observations into a single dataset for each cluster field. Since the
original spectroscopic targets were determined from a separate
catalog, we first cross-matched the spectroscopic catalog with the
optical catalog using a growing annulus technique where the angular
distance $\psi$ was determined using the formula
\[
\psi =
\arccos(\sin(\phi_{s}) \sin(\phi_{p}) + 
	\cos(\phi_{s}) \cos(\phi_{p}) \cos(\theta_{s} - \theta_{p}))
\] 
where the angles ($\theta \equiv$ Declination, $\phi \equiv$ Right
Ascension) have been properly converted into radians, and the
subscript, $s$, refers to spectroscopic target and the subscript, $p$,
refers to photometric object. In all cases, the match-up occurred for
very small angles because spectroscopic targets were selected from the
optical images.

The cross-identification between the infrared and optical catalogs,
however, was slightly more complicated, although it was done in an
identical manner. This was due to the different telescopes and
detectors used during the observations, and the corresponding geometric
distortions present between the two catalogs. Furthermore, the
variation of the spectral energy distributions of the sources in our
sample between the $I$ and $K^{'}$ bands can be quite large, especially
for galaxies at relatively high redshift. As a result, some objects in
each catalog remained unmatched below the angular limit which is
approximately the confusion beam width (approximately five arcseconds).

\section{Empirical Photometric Redshift\label{photoz-sec}}

The technique that we have developed for determining cluster
membership requires redshift estimates for all galaxies in the
field. Although the technique is independent of the actual redshift
estimation technique, we have used the empirical photometric redshift
technique~\citep{brunner99} to generate redshift estimates for all
galaxies in our sample. This is the result of the simplicity of the
empirical technique in determining redshifts for a photometric sample
of galaxies for which an adequate training set exists. Furthermore,
the empirical technique, since it is based only on the data under
investigation, is more robust than alternative techniques to
uncertainties in the spectral energy distributions of galaxies. These
alternative techniques, such as template-based methods, could also be
used, as long as a realistic estimate of the error in the estimated
redshift can be determined.

The empirical photometric redshift technique requires a calibration
sample to determine the coefficients of the empirical relation. For
this work, we have used the spectroscopic data which was acquired as part
of the original observational campaign to determine the cluster
velocity dispersions~\citep{postman98,postman00}. The reliability (\ie
the inverse of the intrinsic error) of a spectroscopic redshift is
generally quite high, with quoted errors often measured in the tenths
of a percent (\ie $\delta < 0.001$). In reality, however, a redshift
is accurate only when the spectral identification is also accurate. As
a result, we restricted the spectroscopic calibrators to include only
objects with redshifts that are determined by at least two spectral
lines, {\it i.e.} a quality of 2 or higher~\citep[for details,
see][]{postman98}. The next step was to restrict the sample to those
objects which were below the $10\%$ photometric error limit, which did
not have bad detection flags (\eg object near edge of frame,
incomplete aperture data), and which were below our high redshift
cut-off of $z = 1.5$.  We also removed one additional galaxy which,
having an $E+A$ spectrum~\citep{dressler83,dressler92,zabludoff96},
skewed the results of our calibration for normal galaxy spectra (which
are the dominant population in our sample).  The final spectroscopic
calibration data consisted of 130 galaxies from five different cluster
fields.

\begin{figure}
\plotone{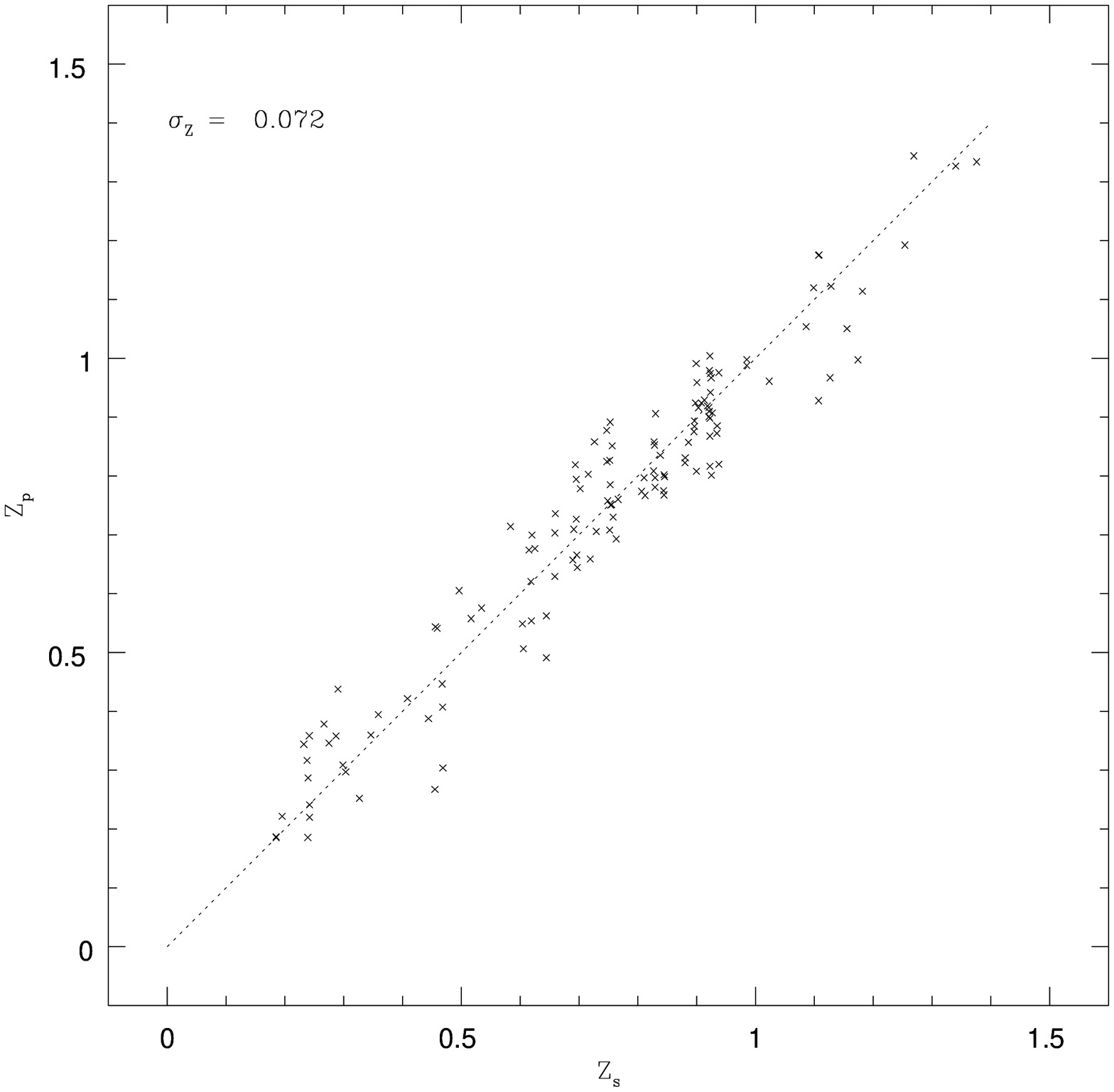}
\caption{The correlation between the photometric 
and spectroscopic redshifts for the entire calibration sample. The
straight line is of unit slope and is not a fit to the actual data.
\label{sigmaz}}
\end{figure}

\begin{figure}
\plotone{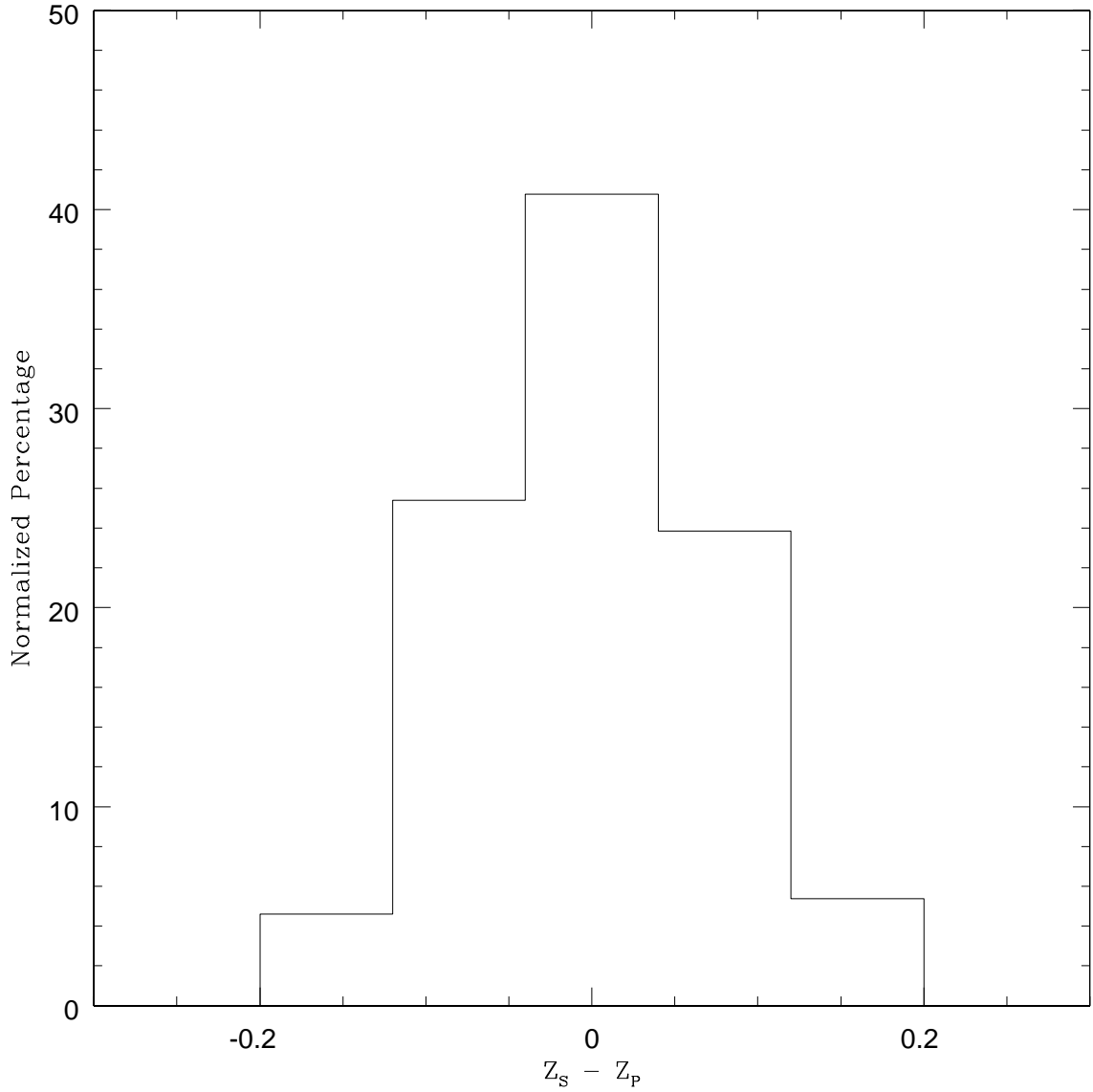}
\caption{A histogram of the residual differences
between photometric and spectroscopic redshifts for the entire
calibration sample, which is approximately Gaussian in shape. The bin
width was chosen to approximate the measured intrinsic dispersion in
our relationship.
\label{histz}}
\end{figure}

With the final calibration sample, we determined a fourth order
polynomial fit to the four magnitudes \BRIK, and the spectroscopic
redshift. Our sample of 130 calibrators is nearly twice the number of
coefficients required (a fourth order polynomial in four variables has
70 coefficients); we are, therefore, properly constraining the degrees
of freedom when determining the coefficients of our polynomial
fit. From Figure~\ref{sigmaz}, there are no obvious systematic
variations in our calculated relationship, which has an intrinsic
dispersion (the standard deviation of the residual differences for the
calibrating sample) of $\sigma_z = 0.072$. In addition, the
distribution of the residuals (see Figure~\ref{histz}) are
approximately Gaussian in shape.

\subsection{Estimating Photometric Redshift Errors}

In order to reliably determine cluster membership, we also need to
have an estimate for the error in our photometric redshift
calculation. While the measurement of the intrinsic dispersion in our
relationship provides one estimate (which is commonly used throughout
the relevant literature, \eg~\citealt{kodama99}), we have also
developed an alternative, extrinsic estimate. Briefly, the intrinsic
error is calculated directly from any available spectroscopic
redshifts; that is, it is simply the dispersion in the derived
relationship between spectroscopic and photometric
redshifts. Extrinsic errors, on the other hand, are calculated in a
Monte-Carlo fashion, accounting for all known measurement
uncertainties.

Following~\cite{brunner99}, we derive extrinsic redshift error
estimates for all galaxies in our dataset using a Monte-Carlo
technique to simulate our uncertainty in both the observables (\ie
flux measurements), as well as our sampling of the galaxy distribution
in the \BRIK\ four-dimensional flux space. Specifically, we use a
bootstrap with replacement algorithm to randomly draw a new set of
calibrating galaxies from the original training set.  This algorithm
allows for duplicate galaxy entries and is designed to emphasize any
incompleteness in the sampling of the true distribution of galaxies in
the four dimensional space \BRIK\ by the calibration
redshifts. Furthermore, the magnitudes of the newly derived
calibrating sample were drawn from a Gaussian probability distribution
function with the mean given by the measured magnitude and sigma by
the magnitude error. This accounts for any uncertainties in the flux
measurements.

This technique was used to generate one hundred different realizations
of the photometric redshift relationship. For each different
realization (all of which form our ensemble), a photometric redshift
was calculated for every galaxy in the photometric catalog.  As a
result, this produces one hundred different redshift estimates for
each galaxy. By applying a quantile cut, we can approximate the
redshift distribution for each individual galaxy by a
Gaussian. Therefore, the error in the photometric redshift for each
object was defined by $\sigma_{z} = (Q5 - Q2) / 2.0$, where $Q5$ and
$Q2$ refer to the fifth and second quantiles of the distribution of
each galaxy's estimated redshift,
respectively~\citep[see][]{brunner99}. Together, our original
photometric redshifts and redshift error estimates completely define
the photometric redshift distribution for each individual galaxy.

Of the two types of redshift error estimates, the intrinsic error is
by far the most commonly used~\citep[\eg][]{kodama99}, as it is the
easiest to calculate. The extrinsic error, however, does a better job
of quantifying the actual error in the redshift estimate of an
individual galaxy, which can be both smaller or considerably larger
than the intrinsically estimated global error~\citep[see Figure 15
in][]{brunner99}. Fundamentally, the extrinsic method provides an
error estimate separately for each individual galaxy, while the
intrinsic method provides only a global estimate.

\section{Cluster Membership\label{pzcluster-sec}}

While the optimal technique for unambiguously determining cluster
membership is to obtain spectroscopic redshifts for all sources in the
field, this process can be prohibitively expensive in telescope time
allocation, especially at relatively high redshifts ($z > 0.5$). As a
result, many previous
studies~\citep[\eg][]{aragon93,lubin96,smail97,dressler97,couch98,stanford98,lubin98a}
have utilized a background contamination estimator in order to
statistically quantify both the number of galaxies which belong to a
given cluster and the morphological fraction of the cluster
members. In addition to possible systematic effects which are implicit
in using techniques of this type, background corrections provide no
information on the likelihood of an individual galaxy being a member
of a cluster.

Photometric redshifts, however, provide a relatively inexpensive
redshift estimate; therefore, they can be used to estimate cluster
redshifts~\citep[\eg][]{gal00} and to determine likely cluster
members~\citep{kodama99,lubin-ociw99}. Previous techniques to select
cluster members based on photometric redshifts~\citep[\eg][]{kodama99}
are based on a simplistic technique for defining galaxy cluster
members. Basically, any galaxy which lies within a redshift shell
centered on the galaxy cluster's redshift (\ie $z_c \pm \delta_z$) are
identified as members. Because photometric redshifts have a
large uncertainty (generally a factor of ten to fifty higher than
spectroscopic redshifts), a large redshift shell (\eg $\delta_z
\approx 0.05$) is used to identify likely cluster members. While
certainly useful, this approach does not provide a reliable means for
identifying line-of-sight contaminating galaxies.

As a result, we have developed an alternative technique which
relies on the probabilistic interpretation of a photometric redshift
to determine cluster membership. We define the probability density
function, $\Phi(z)$, for an individual galaxy's redshift to be a
Gaussian probability distribution function with mean ($\mu$) given by
the estimated photometric redshift and standard deviation ($\sigma$)
defined by the estimated error in the photometric redshift.
\[
\Phi(z) = \frac{1}{\sigma \sqrt{2\pi}} e^{\left(-\frac{(z -
\mu)^2}{2\sigma^2}\right)}
\]
Using this interpretation, we can calculate the probability that a
galaxy has an actual redshift within a given redshift interval.
\begin{eqnarray*}
P(Cluster| z_c, \Delta z, \mu, \sigma) 	
		& = & N \int_{z_{c} - \frac{1}{2} \Delta z}^{z_{c} + \frac{1}{2} \Delta z} 
			\Phi(z) dz \\
		& = & N \int_{z_{c} - \frac{1}{2} \Delta z}^{z_{c} + \frac{1}{2} \Delta z}  
			\frac{1}{\sigma \sqrt{2\pi}} 
			e^{\left(-\frac{(z - \mu)^2}{2\sigma^2}\right)} dz \\
		& = & \frac{N}{2} (\gamma[\frac{1}{2},z_{H}] - \gamma[\frac{1}{2},z_{L}])
\end{eqnarray*}
where $N$ is a suitable normalization factor, $z_c$ is the cluster
redshift, $\Delta z$ is the width in redshift space which you are
sampling, the limits of integration are
\[
z_H = \frac{(z_c + \frac{1}{2} \Delta z) - \mu}{\sqrt{2}\ \sigma}, 
z_L = \frac{(z_c - \frac{1}{2} \Delta z) - \mu}{\sqrt{2}\ \sigma},
\]
and $\gamma$ is the incomplete gamma function. (Of course,
$\gamma[\frac{1}{2},z]$ is also known as the error function,
$erf(z)$).

Within this formalism, the only undeclared quantity is $\sigma$ or,
alternatively, the uncertainty in the estimated redshift for a given
galaxy. In our framework, this value can either be determined from the
intrinsic error in our photometric redshift relation or in a
separately calculated extrinsic error. As a result, this approach is
strongly dependent on the actual technique used to estimate the
photometric redshift error and corresponding redshift error. In this
paper, we have applied this technique using empirically derived
photometric redshifts since we have a well-defined calibration dataset
of spectroscopic redshifts. Empirically defined photometric redshifts
are much less sensitive to uncertainties in the shape and the
evolution with redshift of the spectral energy distribution of
galaxies than competing techniques such as template photometric
redshifts~\citep{myThesis}.

On the other hand, the method we have introduced can easily be adapted
to work with alternative redshift estimation procedures, including
template photometric redshift algorithms, by utilizing the derived
dispersion estimate. The empirical-based techniques do have
shortcomings; most notably, the requirement of a high quality training
set and the fundamental limitation that any resultant analysis is
limited to the same region of flux space delineated by the training
set of calibration redshifts. In our case, the spectroscopic survey
of~\citet{oke98} provides a wonderful training set, which completely
samples the region in flux space which encompasses the vast majority
of cluster galaxies. When this is not the case, template based
photometric redshift estimation methods, which can be applied to data
where there are no calibrators, are required. In this case, however,
extreme care must be used to minimize any systematic effects, such as
morphological variation of the redshift error estimates or template
incompleteness.

By using the calibration galaxies as a training set, we can
empirically determine the optimal probability value which provides the
threshold for cluster membership. This can be tuned to either maximize
the completeness or minimize contamination. In this paper, our sample
is small enough that a simple visual threshold determination is
sufficient; however, there is no reason that a more powerful maximum
likelihood or regression analysis technique could not be employed,
especially for larger datasets (\eg the SDSS). Furthermore, additional
information (such as evolutionary color tracks) can be used to select
likely cluster members which can be used to provide a bootstrap
estimator for the probability threshold.

As a demonstration of this technique, we have applied this procedure
to the data described earlier in this paper. We consider two clusters
from our survey (see Section 2 for more information), the first is Cl
0023+0423 at a redshift of $z = 0.84$ (Postman \etal 1998; Lubin,
Postman \& Oke 1998; Lubin \etal 1998), and the second is Cl 1604+4321
at a redshift of $z = 0.92$~\citep{postman00,lubin00}. These two
clusters provide a useful demonstration because the first cluster is
well sampled by our dataset, while the second cluster pushes the
photometric limits of our data. Since the utility of this technique is
so strongly dependent on the exact method used for estimating the
redshift and redshift error for every galaxy in the field, we have
intentionally relegated a more complete discussion of the application
of this technique to our own data to a later, more targeted
paper~\citep{lubin00}.

\begin{figure}
\plotone{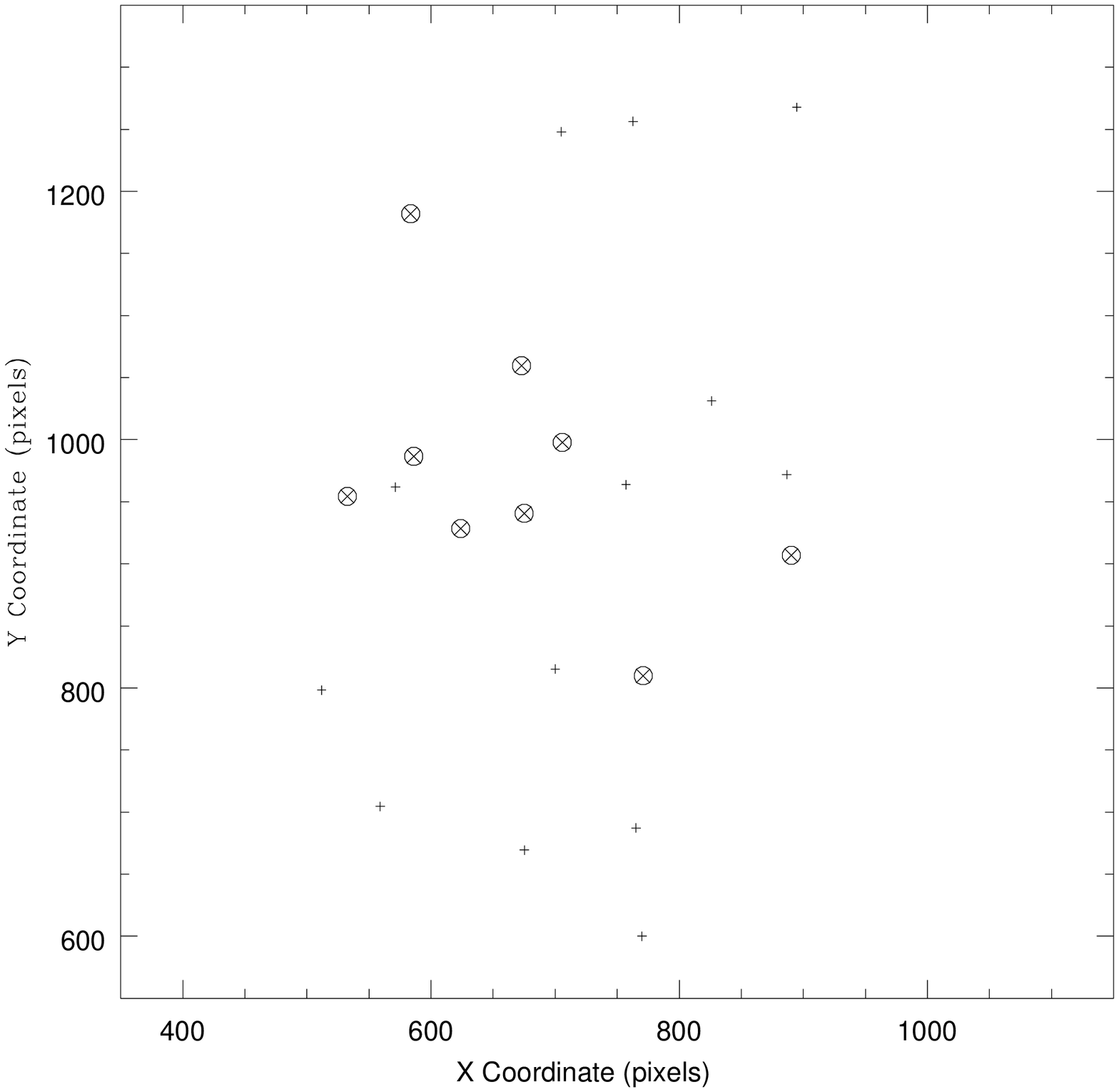}
\caption{Results for the field of Cl 0023+0423. A plot displaying the
locations of calibrating galaxies in the image (the pixel scale is
0.215 arcsec/pixel). The spectroscopically confirmed cluster members
are indicated by $\circ$, photometrically classified cluster members
are indicated by $\times$, and $+$ indicates field galaxies which are
additional calibrators for the photometric redshift relation. With the
probability cut selected for this figure, we have $100 \%$
completeness and $0\%$ contamination.
\label{cl0021s}}
\end{figure}

\begin{figure}
\plotone{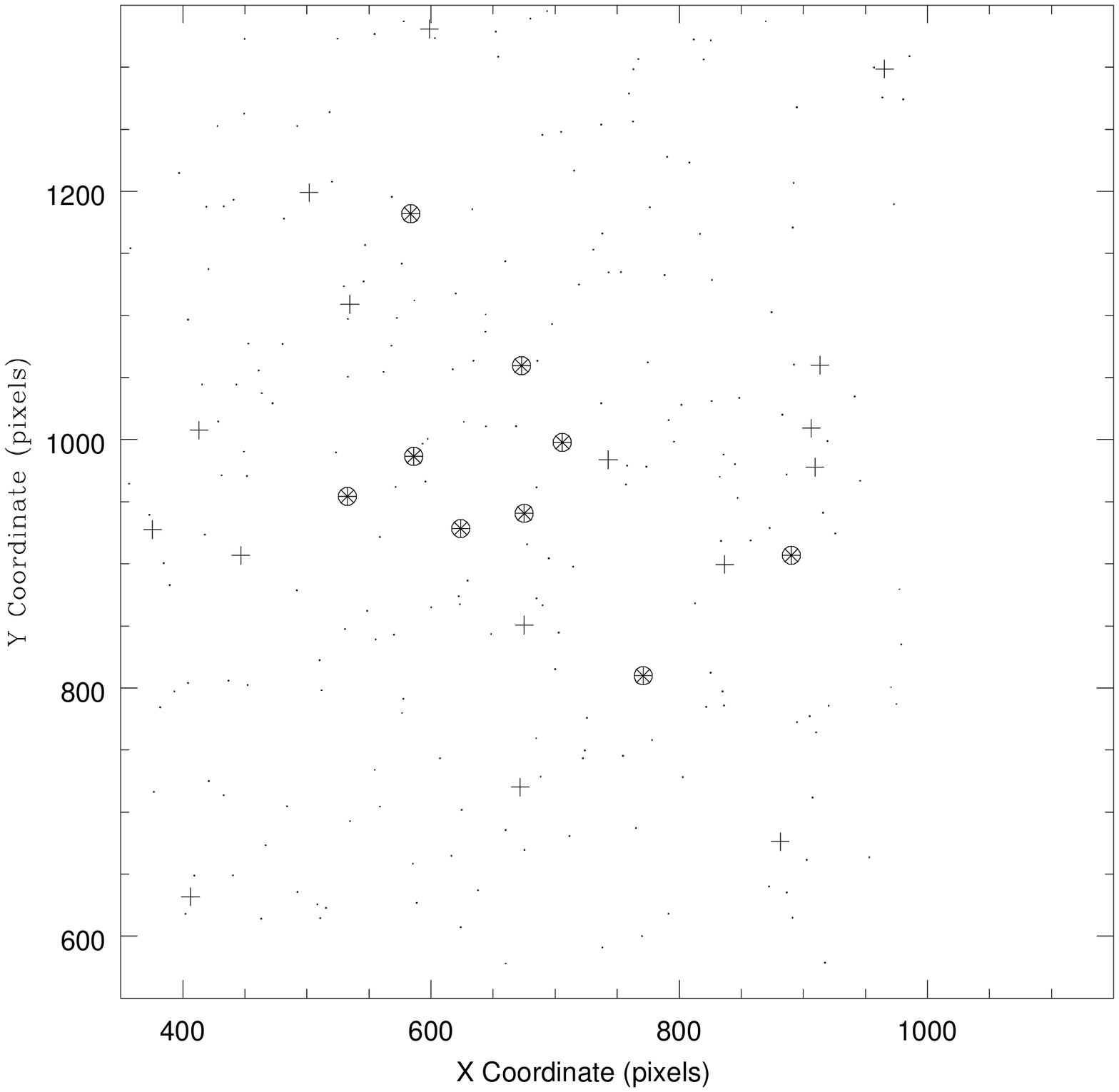}
\caption{Results for the field of Cl 0023+0423. A demonstration of the
application of this technique to the full photometric catalog. Cluster
candidates are indicated as before (both spectroscopic and
photometric), while new cluster candidates are indicated by $+$ and
general field galaxies are indicated by the dots. After correcting for
contamination and incompleteness, the number of cluster members
determined through our photometric redshift technique agrees with the
number predicted by statistically correcting for background counts.
\label{cl0021f}}
\end{figure}

In Figures~\ref{cl0021s} and~\ref{cl0021f}, we display the results of
applying this technique to the spectroscopic targets in the Cl
0023+0423 cluster field, as well as the full photometric sample for
this field. In both cases, we used the intrinsic error of the
photometric redshift relationship as our redshift error. From the
first panel, we clearly do quite well (\ie $100\%$ completeness with
no contamination). Of course, translating this success to the full
sample is not guaranteed; however, as long as our calibration sample
adequately samples the four dimensional flux space occupied by the
galaxies in our full sample (\ie we have a fair and unbiased training
set), there is no reason to expect significant variation from these
results.

\begin{figure}
\plotone{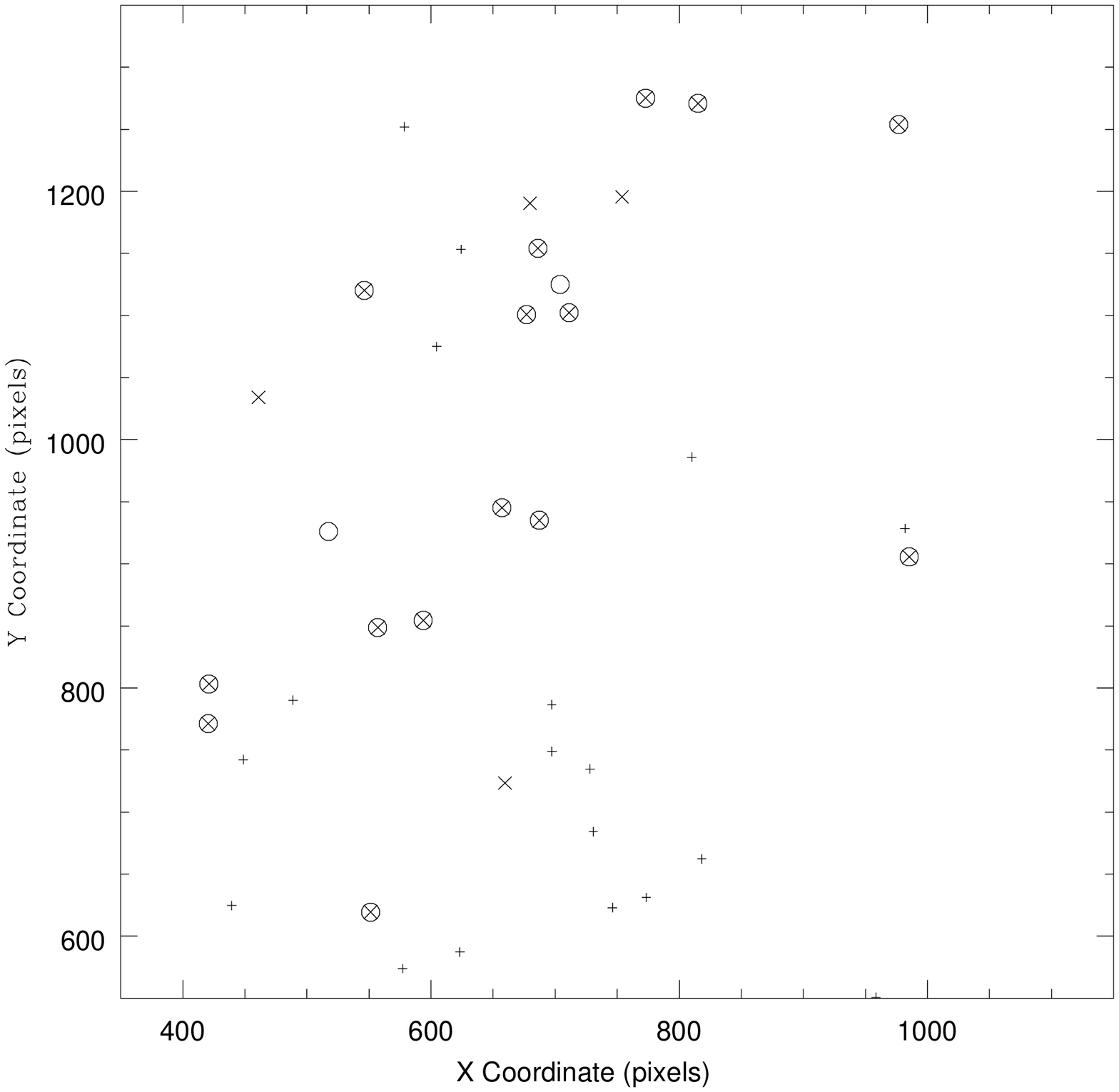}
\caption{Results for the field of Cl 1604+4321. A plot displaying the 
locations of calibrating galaxies in the image (the pixel scale is
0.215 arcsec/pixel). The spectroscopically confirmed cluster members
are indicated by $\circ$, photometrically classified cluster members
are indicated by $\times$, and $+$ indicates field galaxies which are
additional calibrators for the photometric redshift relation. With the
probability cut selected for this figure, we have $\approx 88\%$
completeness and $\approx 21\%$ contamination.
\label{cl1604es}}
\end{figure}

\begin{figure}
\plotone{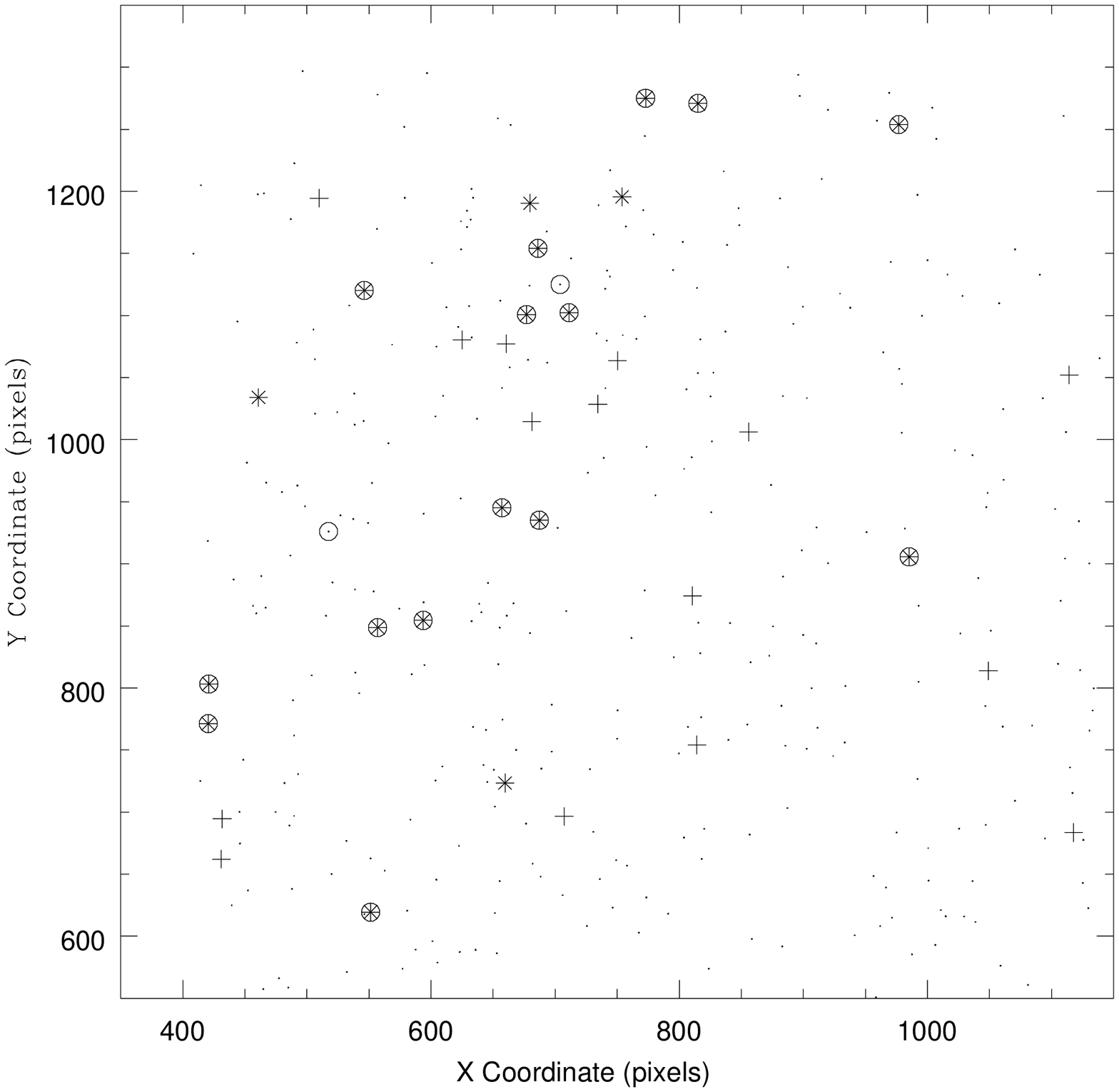}
\caption{Results for the field of Cl 1604+4321.  A demonstration of the 
application of this technique to the full photometric catalog. Cluster
candidates are indicated as before (both spectroscopic and
photometric), while new cluster candidates are indicated by $+$ and
general field galaxies are indicated by the dots. After correcting for
contamination and incompleteness, the number of cluster members
determined through our photometric redshift technique agrees with the
number predicted by statistically correcting for background counts.
\label{cl1604ef}}
\end{figure}

In Figures~\ref{cl1604es} and ~\ref{cl1604ef}, we show the results of
applying this technique to the spectroscopic targets in the Cl
1604+4321 cluster field, as well as the entire photometric sample for
this cluster. This time, however, we have used the extrinsic error
which is calculated independently for each galaxy. By judiciously
selecting the probability threshold, we can maximize the completeness
(which is $\approx 88\%$), while minimizing the contamination (which
is $\approx 21\%$). While we would clearly prefer full completeness
and no contamination, this cluster demonstrates one of the strengths
of our technique, namely the ability to empirically quantify both our
completeness and contamination.

In fact, our technique predicts 37 cluster members within the field of
view of our combined image (see Figure~\ref{cl1604ef}). If we use the
predictions of deep number
counts~\cite[\eg][]{driver95,smail95,abraham96} to estimate the
contamination due to foreground and background galaxies, we estimate
there should be $33\pm14$ cluster members in the image (note the large
uncertainty in this statistical calculation), which is the exact
number that we find after correcting for contamination and
incompleteness [\ie $37 \times (1 - 21 \%) / 88 \% = 33$]. We have
made this comparison with the statistical background correction simply
in order to show that we are finding a reasonable number of cluster
galaxies; however, unlike the background correction technique, we have
explicitly identified the likely cluster members.

\section{Discussion}

In this paper, we have presented a novel technique for the
probabilistic determination of cluster membership based solely on
photometric redshift and redshift error estimates. Our technique can
be tuned to either maximize the completeness of cluster member
identification (\ie identify all actual cluster members), or,
alternatively, to minimize the contamination due to foreground and
background galaxies in the field. Furthermore, this technique provides
a robust estimation for both the completeness and contamination
fractions of identified cluster members as a function of likelihood. A
comparison between our new technique with the traditional technique of
statistical background correction shows remarkable agreement, with the
caveat that our technique actually identifies the galaxies which are
cluster members.

The results in this paper are based on an empirical photometric
redshift estimation, which we have used because (a) we have a reliable
training set from our extensive spectroscopic survey, and (b) these
photometric redshifts are less sensitive to uncertainties in spectral
evolution which can affect other redshift estimation
techniques. Alternative redshift estimation techniques, however, can
easily be incorporated into the algorithm, as all that is needed is a
redshift estimate and a corresponding redshift error estimate. In the
case of template photometric redshifts, the redshift error could be
determined from either the intrinsic dispersion in the photometric
redshift relation, an interpolation over the $\chi^2$ goodness-of-fit
parameter, or even more appropriately through a combination of the
$\chi^2$ parameter and a Monte-Carlo bootstrap approach to mimic
photometric uncertainty and template incompleteness. In this case,
however, caution must be used to minimize any possible systematic
effects which might depend on various fundamental assumptions, such as
the spectral type.

The only free parameter in the entire technique is the probability
threshold used to determine cluster membership. Optimally, as we have
done in this paper, a training set of spectroscopically confirmed
cluster members can be used to empirically set the threshold value.
When this is not the case, alternative techniques, such as
color-selection to identify the early-type sequence at the appropriate
redshift, can be used to identify likely cluster candidates. On the
other hand, the expected number of cluster candidates can be
calculated from the statistical background correction arguments. This
expected number of cluster members can be used as an independent
estimator as we can set the probability threshold value to reproduce
the estimated number of cluster galaxies.

Based on our work, we believe that photometric redshifts are an ideal
way to study the galaxy populations in high-redshift clusters.  For a
large number of the tests of common cluster properties (\eg the
Morphology-Density relationship), this method eliminates the need for
extensive spectroscopic surveys and the uncertainty of estimating the
background contamination. As an additional example of the utility of
this approach, consider the magnitude-limited, spectroscopic survey
used in this analysis. We can approximate the total number of
spectroscopic targets observed by multiplying the number of clusters
used in this analysis (five) by the approximate number of targets per
cluster field (130), implying that a total of 650 objects were
observed spectroscopically. The redshift identification success rate
was nearly $77\%$. Of these redshifts, $50\%$--$85\%$ belong to
non-cluster galaxies. These numbers imply that the spectroscopic
survey was only $12\%$--$38\%$ efficient in identifying cluster
candidates, depending on the richness of the cluster.

In addition, our technique provides the ability to tune the cluster
candidate identification to either maximize the completeness or,
alternatively, to minimize the contamination. As a practical example,
maximizing completeness is essential for any observational program
which is interested in completely sampling the cluster (\eg
spectroscopic targeting). On the other hand, minimizing contamination
is important for selecting a characteristic set of cluster members for
a detailed study of cluster member properties~\citep[\eg the
Morphological fraction or the Morphology-Density
relationship;][]{lubin00}. Furthermore, if a training set of
spectroscopically confirmed galaxies is available, we can actually
characterize the percentages of both the completeness and
contamination that we expect in our final cluster candidate catalog.

Like any photometric redshift based technique, our technique is
limited in its efficacy by the quality of the photometric data used in
the analysis. However, our technique can be applied using a variety of
photometric redshift estimation methods. Therefore, we expect that,
in the near future, we will be able to fully utilize the capabilities
of this technique for cluster research.

\acknowledgments

We are very grateful to Bev Oke and Marc Postman for the enormous time
and effort that they put into obtaining and reducing all of the data
used in this paper.  We also wish to thank the anonymous referee for a
thorough reading of the original paper and many interesting
comments. This research has made use of NASA's Astrophysics Data
System Abstract Service. LML is supported by NASA through Hubble
Fellowship grant HF-01095.01-97A from the Space Telescope Science
Institute, which is operated by the Association of Universities for
Research in Astronomy, Inc., under NASA contract NAS 5-26555.

\newpage

\end{document}